\documentstyle[12pt]{article}

\begin{document}

\vspace{1cm}
{\large
Polar Jahn-Teller Centers and Isotope Effect in Copper Oxide high-$T_c$-
Superconductors.
}

\vspace{0.5cm}
{\it A.S. Moskvin, A.S. Ovchinnikov, Yu.D. Panov, M.A. Sidorov.}

\vspace{0.3cm}
{\small
Department of Theoretical Physics, Ural State University,
620083, Lenin Ave. 51, Ekaterinburg , Russia.
}
\vspace{0.5cm}

{\small
The $CuO_4$-cluster based copper oxides are considered as generalized 
quantum lattice bose-gas or a system of the local singlet bosons moving in a 
lattice of the hole Jahn-Teller centers $[CuO_{4}^{5-}]_{JT}$. The model 
is illustrated by the qualitative and quantitative description of the various 
peculiarities of an isotope shift (IS-) effect in a series of 123-like oxides.
}
\vspace{1cm}

The model approach developed in [1] considers the $CuO_4$-cluster based 
copper oxides as systems unstable with respect to the disproportionation 
reaction
$$
2CuO_{4}^{6-} \rightarrow 
[CuO_{4}^{5-}]_{JT} + [CuO_{4}^{7-}]_{JT}
$$
with the creation of the system of the polar (hole (h-) $CuO_{4}^{5-}$ or 
electron (e-) $CuO_{4}^{7-}$) Jahn-Teller (JT) centers, or a system of 
local bosons (singlet coupled electron pairs) moving in a lattice of singlet-
triplet h-centers. An origin and anomalous properties of h-centres are 
connected with a near degeneracy of the molecular terms $^{1}A_{1g}$ 
(Zhang-Rice singlet), $^{1}E_u$, $^{3}E_u$ for the configurations 
$b_{1g}^2$ and $b_{1g}e_u$ respectively that can create conditions for the 
pseudo-Jahn-Teller (PJT-) effect with the active local displacements modes 
of the $Q_{eu}$-, $Q_{b1g}$- and $Q_{b2g}$-types.

Unusual properties of the $^{1}A_{1g}$, $^{1,3}E_u$-manifold involving 
terms distinguished by the spin multiplicity, parity and orbital degeneracy 
provides unconventional behavior for the h-centers with active interplay of 
various modes.

In general, the PJT-effect in the ($^{1}A_{1g}$, $^{1,3}E_u$)-manifold 
leads to the formation of the four-well adiabatic potential of two symmetry 
types: $E_{u}B_{1g}$ or $E_{u}B_{2g}$. In the first case we have four 
minima with the nonzero local displacements $Q_{eu}\neq 0$, 
$Q_{b1g}\neq 0$ for the hybrid copper-oxygen mode $Q_{eu}$ and the 
purely oxygen mode $Q_{b1g}$. In the second case four wells correspond 
to the nonzero displacements of the $Q_{eu}$- and $Q_{b2g}$-types. In 
both cases we have to deal with the ground state JT-quartets which 
undergoes the tunnel splitting to one doublet and two singlets.

For a weak intermode coupling regime the boson movement is accompanied 
by the induced local structural and spin fluctuations. A microscopic 
mechanism of these fluctuations is connected with the fluctuations of the 
local "bare" parameters such as $A-E$-separations $\Delta_{AE} = 
E(^{1}E_u)-E(^{1}A_{1g})$ and also with the crystalline field 
fluctuations.

One of the most important results of the intermode coupling is an 
appearance of the IS- effect connected with the "vibronic reduction" of the 
boson transfer integral
$$
t_{BB} = t_{BB}^{(0)} \langle \chi_e | \chi_h \rangle^2 
\sim M^{-\alpha}
$$
where $\langle \chi_e | \chi_h \rangle$ is an overlap integral for the 
local oscillatory states with and without local boson, or for the e- and h-
centers respectively.

A vibronic reduction factor 
$K_{eh}=\langle \chi_e | \chi_h \rangle^2$ in general has a very 
complicated structure due to multimode PJT-effect. Normal vibration modes 
in the minima of the adiabatic potential have a hybrid $(b_{1g}-e_{u}x)$ 
and $(b_{2g}-e_{u}y)$ character so that the adiabatic potential cross-
section near minima is a direct sum of the two polarization ellipses, which 
orientation and axis lengths in a complicated manner depend on the 
parameters like $\Delta_{AE}$, vibronic constants, the oxygen and copper 
atomic masses.

In general a functional dependence of the $K_{eh}$ has a form: 
$K_{eh}=N e^{-\gamma}$. For a simplest single-mode PJT-effect the 
$N$-factor does not depend on atomic masses but 
$\gamma \sim \sqrt{m}(\Delta Q)^2$ so that $\alpha \geq 0$ and IS-effect will 
be determined mainly by the distance $\Delta Q$ separating the minimuma 
of the adiabatic potential for e- and h-centers.

For a multimode PJT-effect a situation becomes complicated because of a 
supplementary dependence of the $N$-parameter on atomic masses through the 
orientation and relative magnitudes of the polarization ellipses axis. A 
change in atomic masses results in deformation and turning of the 
corresponding ellipses for the e- and h-centers that provides a change in the 
overlap integral $\langle \chi_e | \chi_h \rangle$. Note that the 
$N$-factor contribution to the IS-parameter $\alpha$ can be both positive and 
negative. A negative IS-effect in our model can be observed practically only 
for the copper atoms.

Real perspectives for a high-$T_c$-superconductivity have to be connected 
with an extremely weak inter-mode coupling regime or with the so called  
{\it optimized systems}. A boson movement in optimized systems does not 
at all accompanied by a significant change in the spin and the local structure 
modes or in other words the charge fluctuations do not result in the spin and 
structure fluctuations. Electron and hole PJT-centers in optimized systems 
have identical oscillatory states with the maximum value of  the "vibronic 
reduction" factor $\langle \chi_e | \chi_h \rangle ^2 = 1$ so the 
IS-effect for these systems is entirely absent.

In practice, the typical special features of the optimized systems are the 
maximally high-$T_c$'s, the minimal width of the superconducting 
transition, the absence of the IS-effect, the minimal value of the baric 
coefficient $dT_c/dp$. Probably, the so called "optimally doped" 
$YBa_{2}Cu_{3}O_{6+x}$ oxide at $x \simeq 0.93$ can be considered as 
one of the real optimized systems.

Note that both $T_c \sim t_{BB} \sim K_{eh}$ and $\alpha_{O,Cu}$ mainly 
depend on the basic PJT-parameters $\Delta_{AE}$ for the e- and h-centers 
($\Delta_e$, $\Delta_h$ respectively) nevertheless the $T_c (\Delta_e, 
\Delta_h)$ and $\alpha (\Delta_e, \Delta_h)$ dependencies are different so 
that there is no simple $\alpha (T_c)$ dependence.

Figure shows a number of calculated model dependencies 
$\alpha (T_c/T_c^{max})$ for different 
"trajectories" (1,2,3) within the 
$(\Delta_e, \Delta_h)$-plane, providing both positive and negative 
$\alpha_{Cu}$-values. Note that we have large positive 
$\alpha_{O}$-values for $\alpha_{Cu}<0$ and moderate positive 
$\alpha_{O}$-values for $\alpha_{Cu}>0$. 
A comparison with experimental data for various 123-
systems $\alpha_{O}$ in: 
$\diamondsuit$ - $YBa_{2}Cu_{3-x}Zn_{x}O_{7-\delta}$ [2],
$\nabla$ - $Y_{0.8-y}Pr_{0.2}Ca_{y}Ba_{2}Cu_{3}O_{7-\delta}$ [2],
$\bigcirc$ - $ Y_{1-x}Pr_{x} Ba_{2}Cu_{3}O_{7-\delta}$ [2],
$\odot$ - $Y Ba_{2-x} La_{x} Cu_{3}O_{7-\delta}$ [3];
$\triangle$ - $\alpha_{Cu}$ in $Y Ba_{2}Cu_{3}O_{7-\delta}$ [4]) 
convincingly evidence the real possibilities for both qualitative and 
quantitative description of the IS-effect within the PJT-centers model.

Above we consider only one of the most important effects ($\Delta_e, 
\Delta_h$-effect) determining the isotope shift. For real systems an IS-effect 
can depend on a number of supplementary factors such as a phase separation 
and percolation phenomena.

The features of the IS-effect in the oxides like $La_{2-x}M_{x}CuO_{4}$ 
are connected with a proximity to $LTO \rightarrow LTT$ (low temperature 
orthorhombic to low temperature thetragonal) transition accompanied by a 
transition to the strong intermode coupling regime with dynamic or static 
phase separation and a suppression of the superconducting state.

\vspace{2cm}
REFERENCES
\vspace{0.5cm}

[1] A.S. Moskvin. Sov.Phys. JETP Lett. {\bf 58}, 342 (1993); A.S. Moskvin, 
Yu.D. Panov. Sov.Phys. JETP {\bf 111}, 644 (1997).

[2] G. Soerensen, S. Gygax. Phys.Rev.B {\bf 51}, 11848 (1995).

[3] H.J. Bornemann, D.E. Morris. Phys.Rev.B {\bf 44}, 5322 (1991).

[4] J.P. Franck, D.D. Lawrie. Physica C {\bf 235-240}, 1503 (1994).

\end{document}